\documentclass[12pt,superscriptaddress,showpacs,showkeys,twocolumn,prb,aps,reprint,a4paper,floatfix]{revtex4-1}

\usepackage{graphicx}
\usepackage{amsmath, amsthm, amssymb, amsfonts}
\usepackage{float}

\usepackage{natbib}
\usepackage{physics}
\usepackage[para,online,flushleft]{threeparttable} 
\usepackage{array}
\usepackage{ragged2e} 

\DeclareUnicodeCharacter{2212}{-}

\makeatletter
\newcommand*{\citenst}[2][]{%
  \begingroup
  \let\NAT@mbox=\mbox
  \let\@cite\NAT@citenum
  \let\NAT@space\NAT@spacechar
  \let\NAT@super@kern\relax
  \renewcommand\NAT@open{[}%
  \renewcommand\NAT@close{]}%
  \citet[#1]{#2}%
  \endgroup
}
\makeatother

\makeatletter
\newcommand*{\citenumns}[2][]{%
  \begingroup
  \let\NAT@mbox=\mbox
  \let\@cite\NAT@citenum
  \let\NAT@space\NAT@spacechar
  \let\NAT@super@kern\relax
  \renewcommand\NAT@open{[}
  \renewcommand\NAT@close{]}%
  \cite[#1]{#2}
  \endgroup
}
\makeatother
\usepackage[caption=false]{subfig}

\begin{document}
\title{Nonlinear characteristics of Ti, Nb, and NbN superconducting resonators for parametric amplifiers}
\author{Songyuan Zhao}
\email{sz311@cam.ac.uk}
\affiliation{Cavendish Laboratory, University of Cambridge, JJ Thomson Avenue, Cambridge CB3 OHE, United Kingdom.}
\author{S. Withington}
\affiliation{Cavendish Laboratory, University of Cambridge, JJ Thomson Avenue, Cambridge CB3 OHE, United Kingdom.}
\affiliation{Department of Physics, University of Oxford, Parks Road, Oxford OX1 3PJ, United Kingdom.}
\author{C. N. Thomas}
\affiliation{Cavendish Laboratory, University of Cambridge, JJ Thomson Avenue, Cambridge CB3 OHE, United Kingdom.}
\date{June 1, 2023}

\begin{abstract}
\noindent  Superconducting resonators and parametric amplifiers are important components in scientific systems such as kinetic inductance detector arrays, frequency-domain multiplexers for other superconducting bolometers, spin-ensemble based memories, and circuit quantum electrodynamics demonstrators. In this paper, we report microwave measurements of superconducting Ti, Nb, and NbN resonators and their use as parametric amplifiers. These half-wave resonators were fabricated under near identical sputtering and lithographic conditions to ensure a like-for-like comparison of material properties. We report a wide range of properties and behaviours in terms of transition temperatures, resistivities, rate-limiting nonlinear response times, nonlinear dissipation, signs of the nonlinear inductances and their dependences on temperature and resonance harmonic. We have successfully operated Nb and NbN resonators as high gain parametric amplifiers, achieving greater than $20\,\mathrm{dB}$ of power amplification. We have shown that for a half-wave resonator, amplification can be realised not only in the fundamental resonance but also in the higher harmonic resonances. Further, for materials with high transition temperatures, e.g. Nb and NbN, amplification can be achieved at $\sim4\,\mathrm{K}$, i.e. a temperature maintained by a pulse tube cooler. Finally, in materials systems that have very fast response times, e.g. NbN, we have found that a cross-harmonic type of amplification can be achieved by placing pump tone in a different resonant mode as the signal and the idler. This wide range of observations will have important implications on the design and application of superconducting parametric amplifiers.
\end{abstract}


\maketitle

\section{Introduction}
Superconducting resonators have important applications in a variety of scientific systems, such as kinetic inductance detector arrays \citenumns{Day_2003}, multiplexed bolometer arrays \citenumns{Dobbs_bolometer_2012,Mates_bolometer_2017}, superconducting qubits \citenumns{Martinis2009}, quantum spin ensembles \citenumns{Wesenberg_2009}, and circuit quantum electrodynamics architectures \citenumns{Wallraff_2004,Mallet_2009}. These systems are, in turn, crucial to progress in fields such as astronomy, quantum computing, and fundamental physics research \citenumns{Jonas_review,Superconducting_Qubits_review, Project8_2009, Hao_2019,Saakyan_2020}. 
In this paper, we report a systematic study on the different nonlinear characteristics of Ti, Nb, and NbN half-wave superconducting resonators, with a special focus on operating these resonators as superconducting parametric amplifiers. 

A superconducting resonator is, in general, a nonlinear device: its dissipative and reactive properties are dependent on the energy stored in the resonator \citenumns{Thomas_2020}. The nonlinearity functions either as a complication, for example in the context of reading out kinetic inductance detector arrays using a strong microwave tone \citenumns{Swenson_2013}, or as an important mechanism that can be exploited to achieve the desired behaviour, for example in the context of DC-biased kinetic inductance detectors to achieve lower frequency detection threshold \citenumns{Zhao_2020}, or to achieve frequency tuneability \citenumns{Vissers_2015,Adamyan_2016_resonator}. An important type of device which makes use of this nonlinearity is the superconducting parametric amplifier.

Superconducting parametric amplifiers are thin-film superconducting devices that achieve amplification through wave-mixing processes induced by the underlying nonlinearity. These devices have been realised in both travelling-wave and resonator geometries. In general, amplifiers based on travelling-wave designs have wider bandwidths, whereas amplifiers based on resonator designs have lower power requirements on the pump tone and are less prone to lithographic defects such as shorts or breaks on the long transmission lines \citenumns{Eom_2012,Shan_2016,zhao2022physics}. Although there are amplifier technologies which rely on the nonlinear Josephson junction inductance \citenumns{Yurke_1988,Movshovich_1990}, in this study, we focus on the type of devices which achieve amplification by exploiting the effect of nonlinear superconducting kinetic inductance. Experimentally, kinetic inductance amplifiers have demonstrated high gain performances of $10-30 \,\mathrm{dB}$ whilst achieving noise performances close the standard quantum limit for linear amplifiers of half a quantum \citenumns{Tholen_2009, Eom_2012, Vissers_2016, Malnou_2020}. By introducing a DC current and operating in degenerate three-wave mixing mode, these amplifiers have also been used to achieve squeezing amplification and the generation of squeezed vacuum states \citenumns{Parker_squeezing_2022}. The remarkable gain and noise characteristics allow kinetic inductance parametric amplifiers to potentially improve the performance of various quantum information and detector systems \citenumns{Ranzani_2018,Zobrist_2019,Vissers_2020}.

Typical analyses of superconducting parametric amplifiers assume an idealised form for the kinetic inductance nonlinearity. Up to second order in the current $I$, the idealised inductance of a superconducting transmission line is described by
\begin{align}
  L = L_0[1+(I/I_*)^2] \, , \label{eq:nonlinear_scale}
\end{align}
where $L$ is the inductance per unit length, $L_0$ is the inductance per unit length in the absence of inductive nonlinearity, and $I_*$ characterises the scale of the nonlinearity. $L$ and $L_0$ contains contributions from both the geometry of lines and the kinetic inductance of the conductors. The nonlinearity affects only the kinetic inductance, and the scaling current as defined relates the size of these changes to the total inductance, rather than just the kinetic inductance. Three assumptions are implicit in Equation~(\ref{eq:nonlinear_scale}): 1) the nonlinear inductance responds instantaneously to changes in the current; 2) the resultant inductance is \textit{increased} by the presence of current; and 3) there is no significant dissipative effect associated with the current. As measurements in this paper show, depending on the choice of material and geometry, the nonlinear behaviour of a parametric amplifier may depart significantly from the above assumptions, often to the detriment of its performance in terms of gain, bandwidth, or added noise.

In this study, we compare the different nonlinear characteristics of Ti, Nb, and NbN superconducting parametric amplifiers based on half-wave resonator geometries. These materials cover a wide range of properties in terms of elemental versus alloy superconductors, resistivities, and transition temperatures. These materials may also have very different underlying nonlinear mechanisms, resulting in different characteristics in terms of powers required to achieve nonlinear wave-mixing, response times of nonlinear inductances, and relative magnitudes of nonlinear dissipation \citenumns{Thomas_2022, Zhao_2022, Adamyan_2016}. Specifically, the overall dissipative characteristics of a resonator may be a complex combination of linear dissipation, nonlinear static dissipation, and nonlinear dynamic dissipation \citenumns{Thomas_2022}. In order to understand the different behaviours of these different materials, a systematic comparison has been performed. To ensure a like-for-like comparison of materials properties and to highlight any differences, all devices in this study were fabricated under close to identical sputtering and lithographic conditions, using the same mask set and the same deposition facilities at the University of Cambridge. Subsequently, the devices were tested using identical microwave measurement systems. Two types of measurements were used to interpret the nonlinear characteristics of a materials. Firstly, we measured sweeps of resonator transmission against frequency at varying power levels. This measurement encapsulates characteristic nonlinear resonator behaviours such as shift in resonant frequency, response bifurcations, and changes in resonance Q-factors \citenumns{Thomas_2020,Swenson_2013}. Secondly, we biased a resonator with a microwave pump tone and measure the resonator transmission using a weak sweep tone. In this case the weak sweep tone is amplified by the strong pump tone and the measurement directly informs us about the amplification characteristics of our resonators. Throughout the measurement result sections, conclusions are drawn from our detailed theoretical analysis on the effects of reactive, dissipative and rate-limited nonlinearity in the context of superconducting resonator parametric amplifiers \citenumns{Thomas_2022}.

Our measurements have led to a collection of important observations relating to the design and operation of superconducting parametric amplifiers: 1) the inductance of Ti shows long characteristic time ($\sim74\,\mathrm{\mu s}$) when responding to a change in current; 2) the overall nonlinearity of Nb and NbN may \textit{decrease} the effective inductance in the fundamental resonant mode but \textit{increase} it in the higher harmonics; similarly, 3) for Nb and NbN, the sign of the nonlinear inductance at $\sim 100\,\mathrm{mK}$ may be different to the sign at $\sim 4\,\mathrm{K}$; 4) Ti displays significant nonlinear dissipation which may prevent the materials from achieving amplification despite significant wave-mixing; 5) for a half-wave resonator, amplification can be achieved not only in the fundamental resonance but also in the higher harmonic resonances; 6) for materials with high transition temperatures, e.g. Nb and NbN, amplification can be achieved at a temperature maintained by a pulse tube cooler, i.e. $\sim4\,\mathrm{K}$; and 7) in material systems that have very fast response times, e.g. NbN, a cross-harmonic type of amplification can be achieved by placing pump tone in a different resonant mode as the signal and the idler. Cross-harmonic amplification is likely to have important practical value as it allows simpler schemes to isolate and extract the signal tone from the pump tone. This is likely to significantly simplify the design of amplifier systems and expand their potential application. In summary, the wide range of observed behaviours for different material systems highlight the need for careful material analysis in the design of superconducting parametric amplifiers.

\section{Fabrication and measurement details}

\begin{figure}[!ht]
\includegraphics[width=8.6cm]{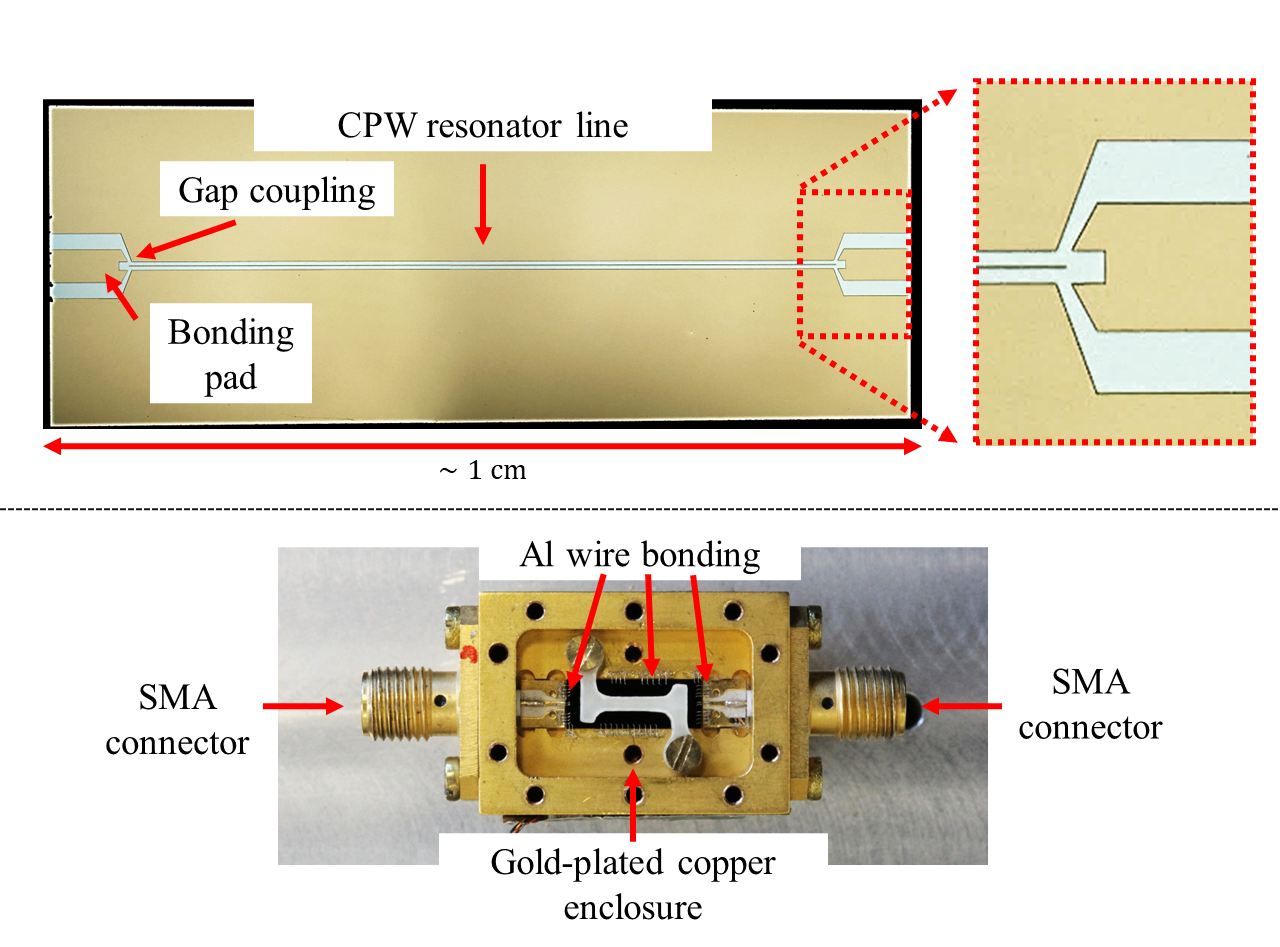}
\caption{\label{fig:resonator_diagram} Top subfigure: an example of the half-wave resonators measured in this study. The resonator comprises a length of coplanar waveguide capacitively coupled in series between two contact pads. Bottom subfigure: gold-plated copper enclosure with SMA connectors for microwave measurement.}
\end{figure}

All devices tested in this study were half-wave resonators based on co-planar waveguide (CPW) geometries, as shown in the top subfigure of Figure~\ref{fig:resonator_diagram}. The CPW conductor width is $5\,\mathrm{\mu m}$, the gap width from conductor to ground is $50\,\mathrm{\mu m}$, and the substrate is Si. These resonators are end coupled through capacitive gap coupling. The coupling Q-factor of the resonator can be controlled by varying the length over which gap coupling occurs. More technical details on the design of these half-wave resonators have been reported in \citenumns{Tess_2021}.

Ti, Nb, and NbN films were deposited onto $50\,\,\rm{mm}$ diameter Si wafers by DC magnetron sputtering at a base pressure of $2\times 10^{-10}\,\,\rm{Torr}$  or below. NbN films were deposited by reactive sputtering a Nb target in the presence of gaseous $\mathrm{N_2}$. All films were deposited at ambient temperature. The film thicknesses were controlled by calibrating and varying the deposition time. All films in this study have thicknesses of $\sim100\,\mathrm{nm}$. The films were patterned using a lift-off process and the same photomask was used for all devices. The wafers were diced into individual half-wave resonators which were bonded using Al wires to a gold-plated copper enclosure with SMA connectors. A photo of the enclosure is shown in the bottom subfigure of Figure~\ref{fig:resonator_diagram}. The enclosure was designed such that its resonance frequencies do not overlap significantly with the resonance frequencies of the half-wave resonators \citenumns{Tess_2021}. The samples were attached to the cold stage of an adiabatic demagnetization refrigerator and the temperature was monitored using a calibrated ruthenium oxide thermometer. Stainless steel coaxial cables were used in the cryogenic system, thermally anchored at $35\,\mathrm{K}$, $4\,\mathrm{K}$, $1\,\mathrm{K}$, and $50\,\mathrm{mK}$ stages. HEMT amplifiers were not used in the signal chain since added-noise measurement of parametric amplifiers was not an aim of this study. Experiments were performed at $\sim100\,\mathrm{mK}$ and $\sim4\,\mathrm{K}$ to characterise properties and behaviours of the materials at different temperatures.

All measurements were taken using a vector network analyser (VNA). In the context of half-wave resonator based amplifiers, power transmission is an important quantity. We define pre-normalisation power transmission $T_\mathrm{out,0}$ as the ratio between the power received at the VNA $P_\mathrm{out, VNA}$ and the power delivered by the VNA into the cryogenic system $P_\mathrm{in, VNA}$. It includes both the effect of the device under test as well as the attenuation due to coaxial cables in the cryostat, which was around $-30\,\mathrm{dB}$ at $5\,\mathrm{GHz}$ in our case.
Using a VNA, $T_\mathrm{out,0}$ in units of \textrm{dB} is calculated using $T_\mathrm{out,0}=20\log_{10}|S_{21}| \,\, \mathrm{dB},$
where $S_{21}$ is the forward scattering parameters from the output port of the VNA to the input port. In order to better compare the effect of the underlying nonlinearity, we normalise $T_\mathrm{out,0}$ to obtain $T_\mathrm{out}$ such that the maximum of $T_\mathrm{out}$, on resonance, is unity ($0\,\mathrm{dB}$) at the lowest power setting.

In this study we focus on two main types of measurements: transmission measurements (without a pump tone) and amplification measurements (in the presence of a pump tone). In the transmission measurements, resonator power transmission $T_\mathrm{out}$ as a function frequency $f$ was measured at varying input power levels $P_\mathrm{in}$, measured at the input of the cryostat. For a fixed $P_\mathrm{in}$, the maximum of a transmission sweep occurs at $f_\mathrm{max}$ with transmission $T_\mathrm{out, max}$. Key characteristics of the underlying nonlinearity can be deduced from analysing the transmission measurements \citenumns{Thomas_2020,Swenson_2013}. Shifts in the resonant frequency $f_\mathrm{max}$ against $P_\mathrm{in}$ can be used to understand the functional forms of the nonlinear inductances. Changes in transmission levels and resonator Q-factors can be used to infer the nonlinear dissipation characteristics. At high sweep power levels, the scattering parameters of a superconducting resonator become hysteretic with respect to sweep directions \citenumns{Swenson_2013,Thomas_2020}. This nonlinear phenomenon is commonly referred to as response bifurcation and it is closely linked with the potential of a resonator to achieve high gain. The power density at which bifurcation occurs can be used as a characteristic power density requirement for operating a resonator as an amplifier \citenumns{Thomas_2022}.

To obtain amplification measurements, we biased a resonator with a microwave pump tone at power level $P_\mathrm{pump}$ and measured the resonator transmission using a weak sweep tone as the signal. The signal tone and the pump tone were combined at the input of the cryostat using a multiband two-way power combiner. Care has been taken to ensure that, apart from the resonator amplifier, the measurement system was linear and did not result in unwanted wave-mixing. The weak sweep tone was amplified by the strong pump tone through wave-mixing induced by the nonlinearity of the half-wave resonator. The power of the weak sweep tone was at least $50\,\mathrm{dB}$ smaller than the power of the pump tone in all experiments. This small signal power was chosen to ensure that the signal tone did not result in gain saturation \citenumns{Pozar_2011,Jurkus_1960,Songyuan_2019_paramp,Saturation_JPA}. We have experimentally confirmed that, within $6\,\mathrm{dB}$ around the chosen power levels, the output signal power was linear with respect to the input signal power. For a resonator amplifier, finding a suitable pump frequency and power combination as an operating point is a difficult problem. Guided by our previous theoretical analysis \citenumns{Thomas_2022}, we were able to achieve stable, high-gain amplification by placing the pump close to the bifurcation point of the resonator. Since, for a half-wave resonator, the transmission on resonance at low sweep power is close to unity \citenumns{Thomas_2020}, the gain as a function of frequency and pump power can be equated to $T_\mathrm{out}$. To distinguish amplification measurements from transmission measurements, we denote the gain as $G(f, P_\mathrm{pump})=T_\mathrm{out}(f, P_\mathrm{pump})$. Care had been taken to phase-lock the network analyser, i.e. source of the signal sweep tone, with the signal generator, i.e. source of the strong pump tone, using a single $10\,\mathrm{MHz}$ reference. The frequency steps and intermediate frequencies bandwidths of the network analyser had been carefully chosen such that the pump tone appears in a single output frequency bin in all sweeps.

\begin{figure}[!ht]
\includegraphics[width=8.6cm]{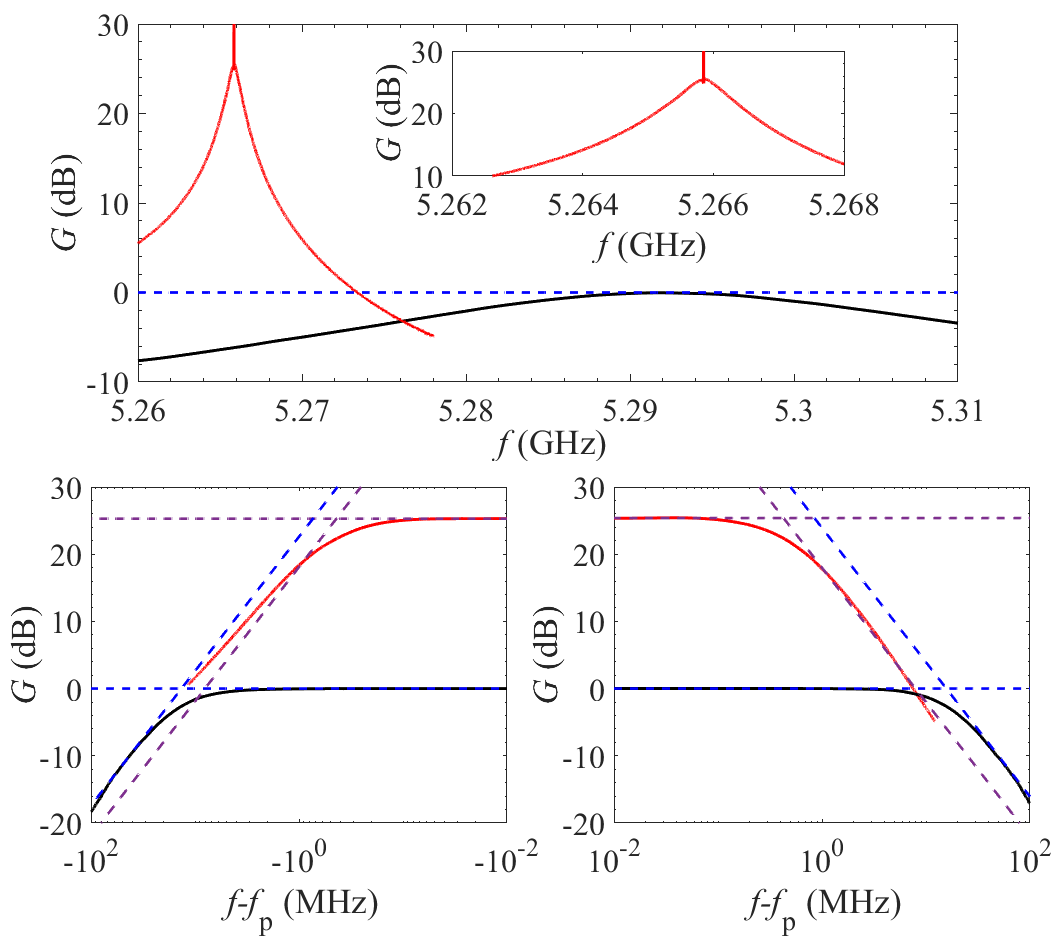}
\caption{\label{fig:all_compiled} Example of amplification measurement of a NbN resonator amplifier. Black line: response of the resonance in the absence of the pump tone; blue dashed line: reference line of unity gain, i.e. $0\,\mathrm{dB}$; red line: response of the resonance in the presence of the pump tone with the resonator now acting as an amplifier. The sharp spike at $\sim5.266\,\mathrm{GHz}$ is the strong pump tone. Top subfigure: the resonance and amplification are plotted against the absolute frequency on a linear scale; bottom subfigures: the resonance and amplification are plotted against the relative frequency $f-f_\mathrm{p}$ on logarithmic scales. For the black lines, $f_\mathrm{p}$ indicates the resonant frequency in the absence of the pump tone; for the red lines, $f_\mathrm{p}$ indicates the frequency of the pump tone. For this particular amplification measurement, the power of the pump tone at the input of the cryostat was $3\,\mathrm{dBm}$, and it was placed in the second harmonic of the resonator. In the bottom subfigures, the horizontal dotted lines indicate the height of the transmission or amplification plateaus; the inclined dotted lines indicate the effect of first-order roll-off, i.e. $20\,\mathrm{dB}$ per decade; the dotted lines intersect at frequencies where the transmission/amplification has decreased by $3\,\mathrm{dB}$. }
\end{figure}

Figure~\ref{fig:all_compiled} shows an example of amplification measurement of a NbN resonator amplifier. This particular NbN resonator was designed to have a low Q-factor (compared to the other NbN resonators shown in later sections of this study) in order to increase its bandwidth. The operating point of this amplifier has been carefully optimised to achieve high gain. The black lines indicate the response of the resonance in the absence of the pump tone, whereas the red lines indicate the response of the resonance in the presence of a strong pump tone, with the resonator now acting as an amplifier. The power of the pump tone at the input of the cryostat is $3\,\mathrm{dBm}$, which translates to a power of around $-12\,\mathrm{dBm}$ at the input to the parametric amplifier. In the absence of the pump tone, the resonant frequency of the resonator is $5.292\,\mathrm{GHz}$; when the pump tone is introduced, the resonant frequency shifts due to the underlying nonlinear mechanisms. High gain can be achieved when the pump tone is placed close to the peak of the shifted underlying resonance \citenumns{Thomas_2022}, in this instance at $5.266\,\mathrm{GHz}$. The sharp spike in the inset of Figure~\ref{fig:all_compiled} shows the pump tone clearly against the smooth amplification profile of the resonator. We have chosen not to remove the pump tone in our figures as it is helpful in visually indicating the pump frequency. In the bottom subfigures, the resonance and amplification are plotted against the relative frequency $f-f_\mathrm{p}$ on logarithmic scales. For the black resonance lines, $f_\mathrm{p}$ indicates the resonant frequency in the absence of the pump tone; for the red amplification lines, $f_\mathrm{p}$ indicates the frequency of the pump tone. Linear frequency plots and logarithmic frequency Bode plots are different ways of displaying the same experimental data. Bode plots of this type are commonly used in amplifier analyses to straightforwardly evaluate maximum gains, bandwidths, and flatness of the mid-frequency ranges. The horizontal dotted lines in the Bode plots indicate the height of the transmission or amplification plateaus whereas the inclined dotted lines indicate the effect of first-order roll-off, i.e. $20\,\mathrm{dB}$ per decade. The two sets of dotted lines intersect at frequencies where the transmission or amplification decrease by $3\,\mathrm{dB}$. As seen in the figure, the first-order roll-off provides a good description of device behaviour for both the resonator system as well as the amplifier system, in agreement with predictions from previous theoretical studies \citenumns{Thomas_2020,Thomas_2022}.

It is practically significant that high gains of greater than $20\,\mathrm{dB}$ can be achieved at a relatively low pump power of $-12\,\mathrm{dBm}$ for our resonator amplifiers. In comparison, the travelling-wave amplifier described in \citenumns{Eom_2012} requires a pump power of $-8\,\mathrm{dBm}$ to achieve $\sim 10\,\mathrm{dB}$ of gain. The difference in power \textit{density} is even greater between the two devices since the travelling-wave amplifier has thickness of $35\,\mathrm{nm}$ and CPW conductor width of $1\,\mathrm{\mu m}$ whereas the resonator amplifiers in this study have thicknesses of $100\,\mathrm{nm}$ and CPW conductor widths of $5\,\mathrm{\mu m}$. This difference arises because, for a given power incident on a resonator, the energy stored on the resonator is enhanced by its Q-factor \citenumns{Thomas_2020}; hence the energy required to achieve nonlinear mixing can be satisfied at a lower incident power. This reduction in power requirement is an important practical advantage of resonator based parametric amplifiers in narrow-band applications.

\section{Titanium measurements}
\begin{figure}[!ht]
\includegraphics[width=8.6cm]{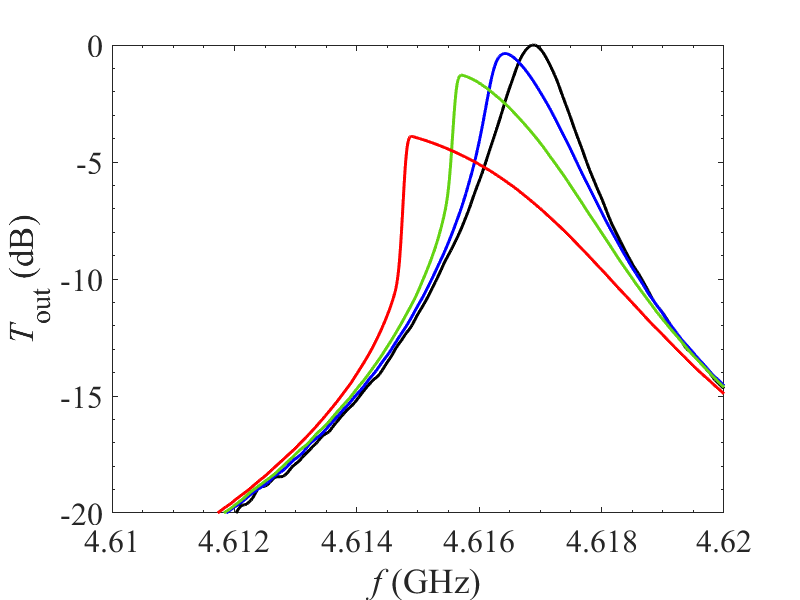}
\caption{\label{fig:Ti_Bifurcation} Measurements of transmission $T_\mathrm{out}$ against frequency $f$ for a Ti half-wave resonator at $0.1\,\mathrm{K}$ for varying levels of sweep power $P_\mathrm{in}$. The sweep direction is from low frequency to high frequency. Black line: $P_\mathrm{in}=-70\,\mathrm{dBm}$; blue line: $P_\mathrm{in}=-60\,\mathrm{dBm}$; green line: $P_\mathrm{in}=-55\,\mathrm{dBm}$; red line: $P_\mathrm{in}=-50\,\mathrm{dBm}$.}
\end{figure}

The Ti films tested in this study have transition temperatures $T_\mathrm{c}=0.55\,\mathrm{K}$ and resistivities $\rho=17\,\mathrm{\mu\Omega\,cm}$. Figure~\ref{fig:Ti_Bifurcation} shows transmission measurement of a Ti half-wave resonator at $0.1\,\mathrm{K}$ for varying levels of sweep power $P_\mathrm{in}$ from $-70\,\mathrm{dBm}$ to $-50\,\mathrm{dBm}$, as measured from the input of the cryostat. The direction of these sweeps is from low frequency to high frequency. As seen in the figure, the transmission of the resonator distorts as the power of the sweep tone increases. In most publications, the shapes of the transmission measurement are explained by shifting of the underlying resonance in response to the presence of the strong sweep tone (which is also known as the readout generator) \citenumns{Swenson_2013}. A characteristic behaviour of nonlinear resonator response is the abrupt change in $T_\mathrm{out}$ at some bifurcation frequency when swept at high levels of $P_\mathrm{in}$. In Figure~\ref{fig:Ti_Bifurcation}, the green line ($P_\mathrm{in}=-55\,\mathrm{dBm}$) shows the transmission sweep at the onset of bifurcation; and the red line ($P_\mathrm{in}=-50\,\mathrm{dBm}$) shows the transmission sweep well into bifurcation.

\begin{figure}[!ht]
\includegraphics[width=8.6cm]{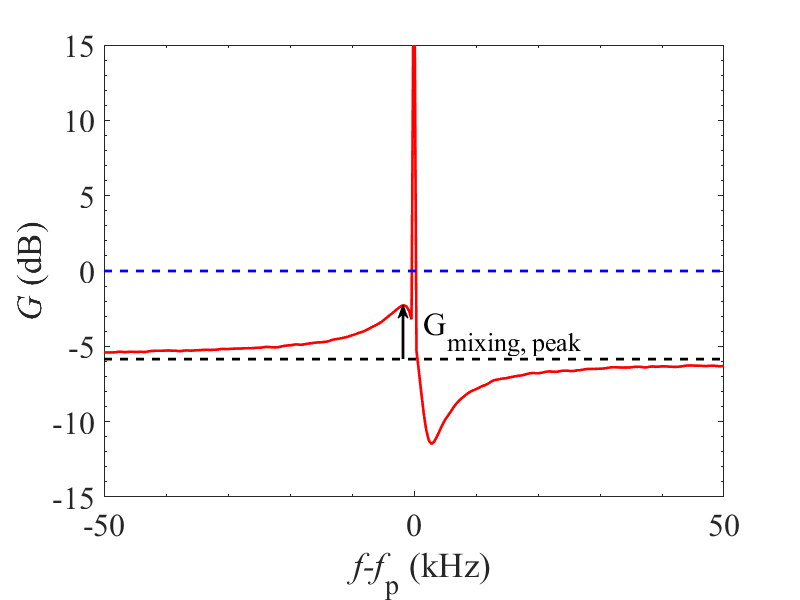}
\caption{\label{fig:Ti_amplification} Measurements of power gain $G$ against frequency $f-f_\mathrm{p}$ for a Ti half-wave resonator, where $f_\mathrm{p}$ is the frequency of the pump tone. The sharp spike at $f-f_\mathrm{p}=0\,\mathrm{kHz}$ is the strong pump tone. The power of the pump tone at the input of the cryostat $P_\mathrm{pump}$ is $-50\,\mathrm{dBm}$ and it is placed at frequency close to the peak of the resonance. }
\end{figure}

The theory of resonator amplifiers predicts that strong wave-mixing behaviour can be obtained by setting the pump tone at power close to that needed to achieve bifurcation \citenumns{Thomas_2022}. Figure~\ref{fig:Ti_amplification} shows amplification measurement of the same Ti resonator using a pump with $P_\mathrm{pump}=-50\,\mathrm{dBm}$, as measured at the input of the cryostat, at frequency close to the peak of the shifted underlying resonance. As seen in the figure, focusing on the vertical axis, the gain obtained from wave-mixing is insufficient to overcome the increase in dissipation, even at frequencies very close to the pump. As a result, the overall gain is less than unity. The effect of the underlying nonlinearity contains both significant reactive effect as well as significant dissipative effect. This points to the importance of including nonlinear dissipation in the analysis of nonlinear resonators, for example through a reactive-dissipative angle characterising the nonlinearity \citenumns{Thomas_2022}. Previous studies of Ti in the context of travelling-wave amplifiers suggest that the ratios of reactive nonlinear effect to dissipative nonlinear effect are in agreement with predictions from the non-equilibrium quasiparticle heating nonlinearity \citenumns{Zhao_2022}. Further, focusing on the horizontal axis of Figure~\ref{fig:Ti_amplification}, the effect of wave-mixing decays by $1\,\mathrm{dB}$ across a bandwidth of $10\,\mathrm{kHz}$ whereas the underlying resonance decays by $1\,\mathrm{dB}$ across a bandwidth of $0.6\,\mathrm{MHz}$. In the absence of rate-limiting mechanisms, the bandwidth of wave-mixing $\Delta f_\mathrm{mixing}$ can be approximated by $\Delta f_\mathrm{mixing}\sim\Delta f_\mathrm{res}/\sqrt{G_\mathrm{mixing, peak}}$, where $\Delta f_\mathrm{res}$ is the bandwidth of the resonance and $G_\mathrm{mixing, peak}$ is the peak power gain from the wave-mixing process \citenumns{Thomas_2022}. For the data in Figure~\ref{fig:Ti_amplification}, $G_\mathrm{mixing, peak}\sim3\,\mathrm{dB}$; in the absence of rate-limiting mechanisms, $\Delta f_\mathrm{mixing}$ should be of the order of $0.3\,\mathrm{MHz}$, much wider compared to the measured value of $10\,\mathrm{kHz}$.  Our measurement thus strongly suggests that the underlying nonlinearity is rate-limited. The amplification response in Figure~\ref{fig:Ti_amplification} also contains a characteristic post-gain dip at $f-f_\mathrm{p}\sim 3 \,\mathrm{kHz}$. This feature results from the phase effect of a rate-limited nonlinear response \citenumns{Zhao_2022}, and it is fully predicted by our model in \citenumns{Thomas_2022}. Using the model of nonlinearity detailed in \citenumns{Thomas_2022}, the underlying nonlinearity of the Ti resonator shown in Figure~\ref{fig:Ti_amplification} has a best-fit response time of $74\,\mathrm{\mu s}$. Similar observations of rate-limited nonlinearities have been made on superconducting Al in the context of travelling-wave amplifiers \citenumns{Zhao_2022}.

\begin{figure}[!ht]
\includegraphics[width=8.6cm]{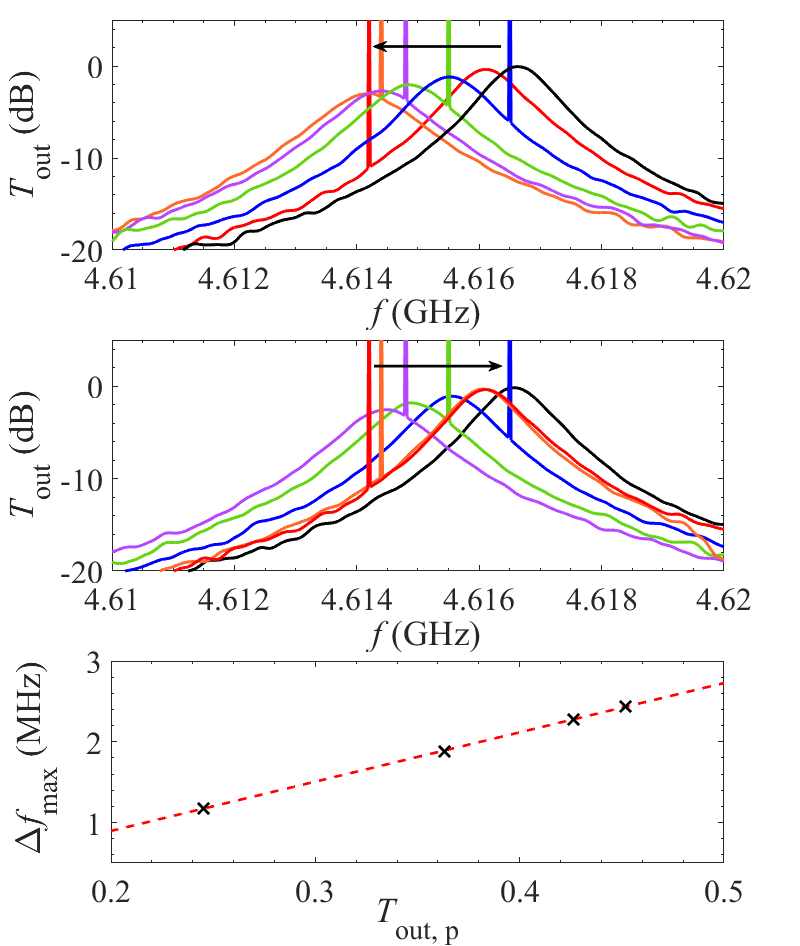}
\caption{\label{fig:Ti_Envelope_compiled} Measurements of transmission $T_\mathrm{out}$ against frequency $f$ for a Ti half-wave resonator in the presence of a pump tone with $P_\mathrm{pump} = -50\,\mathrm{dBm}$ at the input of the cryostat. The black line indicates the resonance in the absence of the pump tone. The pump tone is indicated by the sharp spikes. Top subfigure: the pump tone moves from higher frequencies to lower frequencies; middle subfigure: the pump tone moves from lower frequencies to high frequencies. Bottom subfigure shows frequency shift of resonance $\Delta f_\mathrm{max}$ against power transmission at pump $T_\mathrm{out, p}$: cross markers indicate data points extracted from the top subfigure; dotted line shows the result of fitting with the model $\Delta f_\mathrm{max} \propto T_\mathrm{out, p}$, giving a R-squared value of $0.9998$.}
\end{figure}

Although the distortion of the transmission curve is well understood in terms of shifting of the underlying resonance \citenumns{Swenson_2013}, it is extremely difficult to experimentally measure the latter effect: for most material systems, a measurement of the resonance transmission in the presence of a strong pump tone will also measure the effect of nonlinear wave-mixing \textit{between} the pump tone and the signal sweep tone. In other words, the two effects resulting from a strong pump tone, i.e. resonance shifting and wave-mixing, cannot be separated. For Ti resonators, however, the effect of wave-mixing is restricted to a narrow band of $\sim10\,\mathrm{kHz}$. Thus a transmission sweep of the shifted underlying resonance can be achieved across almost the entire resonance bandwidth without distortions from wave-mixing effects. Figure~\ref{fig:Ti_Envelope_compiled} shows changes in the underlying resonance for a $100\,\mathrm{nm}$ Ti half-wave resonator. The top subfigure shows the pump tone moves from higher frequencies to lower frequencies. As described by Swenson \textit{et al.} \citenumns{Swenson_2013}, the pump chases the peak of the resonance, increasing the detuning of the shifted resonance with respect to the original resonance in the process. At a threshold bifurcation frequency, a positive feedback occurs and forces the resonance to rapidly switch back to a non-energised state. The middle subfigure of Figure~\ref{fig:Ti_Envelope_compiled} shows the same change in the underlying resonance, but the pump tone moves from lower frequencies to higher frequencies. Here the resonance remains in the non-energised state before the detuning of the pump tone is small enough to force the resonance to switch into an energised state. Comparing the orange lines in both figures, we observe that the resonance remains in the energised state for a larger range of frequencies when the sweeping direction is from higher frequencies to lower frequencies. These observations directly confirm the predictions from the analysis conducted by Swenson \textit{et al.} \citenumns{Swenson_2013}. The bottom subfigure of Figure~\ref{fig:Ti_Envelope_compiled} shows frequency shift of resonance $\Delta f_\mathrm{max}$ against power transmission at pump frequency $T_\mathrm{out, p}$. The cross markers indicate data points extracted from the top subfigure while the dotted line shows the result of fitting with the model $\Delta f_\mathrm{max}\propto T_\mathrm{out, p}$. The best-fit line has a high R-squared coefficient of determination of $0.9998$, indicating that the model well-describes the changes in resonance frequency.  Since $\Delta f_\mathrm{max}$ is proportional to the nonlinear inductance and $T_\mathrm{out, p}$ is proportional to the power flow into the nonlinear system \citenumns{Thomas_2022}, the model confirms that, for Ti, the nonlinear inductance depends linearly on the power applied to the resonator.

\section{Niobium measurements}
\begin{figure}[!ht]
\includegraphics[width=8.6cm]{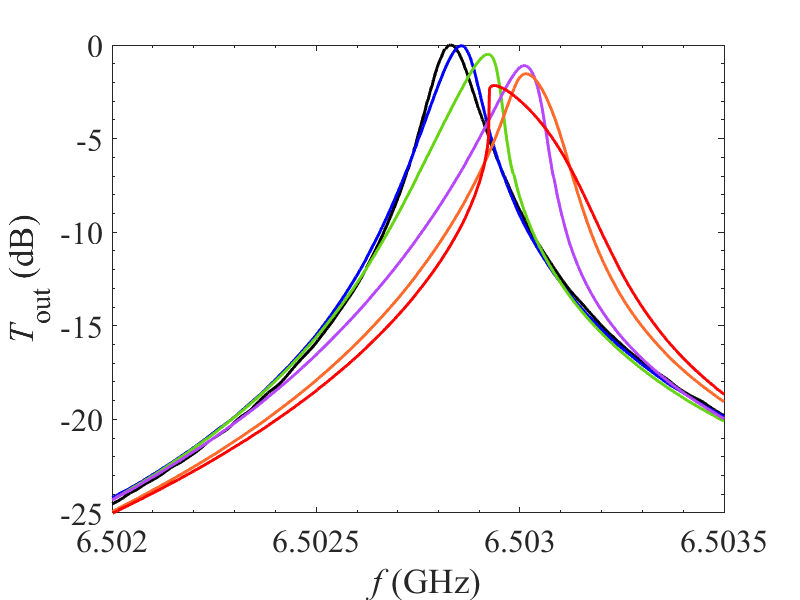}
\caption{\label{fig:Nb_Bifurcation_cold} Measurements of transmission $T_\mathrm{out}$ against frequency $f$ for a Nb half-wave resonator at $0.1\,\mathrm{K}$ for varying levels of sweep power $P_\mathrm{in}$. The sweep direction is from low frequency to high frequency. Black line: $P_\mathrm{in}=-40\,\mathrm{dBm}$; blue line: $P_\mathrm{in}=-30\,\mathrm{dBm}$; green line: $P_\mathrm{in}=-20\,\mathrm{dBm}$; purple line: $P_\mathrm{in}=-10\,\mathrm{dBm}$; orange line: $P_\mathrm{in}=-5\,\mathrm{dBm}$; red line: $P_\mathrm{in}=0\,\mathrm{dBm}$.}
\end{figure}

The Nb films tested in this study have transition temperatures $T_\mathrm{c}=9.2\,\mathrm{K}$ and resistivities $\rho=8.8\,\mathrm{\mu\Omega\,cm}$. Figure~\ref{fig:Nb_Bifurcation_cold} shows transmission measurement of a Nb half-wave resonator at $0.1\,\mathrm{K}$ for varying levels of sweep power $P_\mathrm{in}$ from $-40\,\mathrm{dBm}$ to $0\,\mathrm{dBm}$, as measured at the input of the cryostat. The direction of these sweeps is from low frequency to high frequency. The temperature remained stable within noise levels at every stage of the cryogenic system for all sweep powers. As seen, the transmission of the resonator distorts in response to increasing power of sweep tones. Here we notice contrasting behaviour at low and high sweep powers: at low sweep powers of $-40\,\mathrm{dBm}$ to $-10\,\mathrm{dBm}$, the resonant frequency increases as the sweep power is increased; at high sweep powers of $-10\,\mathrm{dBm}$ to $0\,\mathrm{dBm}$, the resonant frequency decreases as the sweep power is increased. This contrasting behaviour suggests that there might be multiple sources of nonlinearity present in Nb resonators, each dominant at a different power regime. In particular, the increase of resonant frequency with sweep power cannot explained by existing theories of superconducting nonlinear mechanisms prominent in the literature, such as two-level system nonlinearity, non-equilibrium heating nonlinearity of the quasiparticle system, and equilibrium supercurrent nonlinearity \citenumns{Zhao_2022}. Similar observations of increase in resonant frequency with sweep power have been reported in \citenumns{Tholen_2007,Visser_2014} for Nb and Al resonators. The cause of this increase in resonant frequency remains uncertain.

\begin{figure}[!ht]
\includegraphics[width=8.6cm]{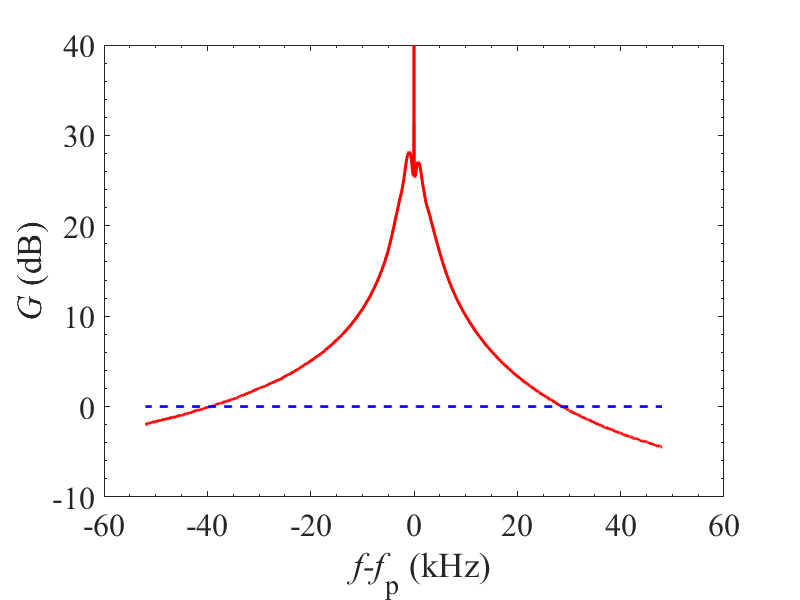}
\caption{\label{fig:Nb_amplification_cold} Measurements of power gain $G$ against frequency $f-f_\mathrm{p}$ for a Nb half-wave resonator at $0.1\,\mathrm{K}$. The sharp spike at $f-f_\mathrm{p}=0\,\mathrm{kHz}$ is the strong pump tone. The power of the pump tone at the input of the cryostat $P_\mathrm{pump}$ is $-0.5\,\mathrm{dBm}$ and it is placed at frequency close to the peak of the resonance.}
\end{figure}

Figure~\ref{fig:Nb_amplification_cold} shows an amplification measurement on the same Nb resonator using a pump with $P_\mathrm{pump}=-0.5\,\mathrm{dBm}$ at frequency close to the peak of the resonance. The amplification resulting from nonlinear wave-mixing is able to overcome any attenuation resulting from nonlinear dissipation. As a result, high gains can be achieved. Gains greater than $20\,\mathrm{dB}$ can be achieved over a range of $10\,\mathrm{kHz}$, and gains greater than $10\,\mathrm{dB}$ can be achieved over a range of $20\,\mathrm{kHz}$. This stands in contrast with the nonlinear behaviour of the Ti resonator shown in Figure~\ref{fig:Ti_amplification}, where the nonlinear effects combine to result in net-attenuation instead of amplification. We also observe here that the amplification profile of the Nb resonator is limited by the bandwidth of the underlying resonance, and not by another rate-limiting mechanism. This is also in contrast with the behaviour of the Ti resonator, where the nonlinear wave-mixing is observed within a narrow-band on top of the broad-band resonance. The result here does not imply that the nonlinear mechanisms of Nb resonators are not rate-limited; instead, it only suggests that these mechanisms not rate-limited at bandwidths of $\sim100\,\mathrm{kHz}$. Although the shift in resonant frequency of the Nb resonator cannot be explained by existing theories, its amplification behaviour shown in Figure~\ref{fig:Nb_amplification_cold} can still be adequately modelled by \citenumns{Thomas_2022}, provided that the sign of the nonlinear inductance is flipped numerically.

\begin{figure}[!ht]
\includegraphics[width=8.6cm]{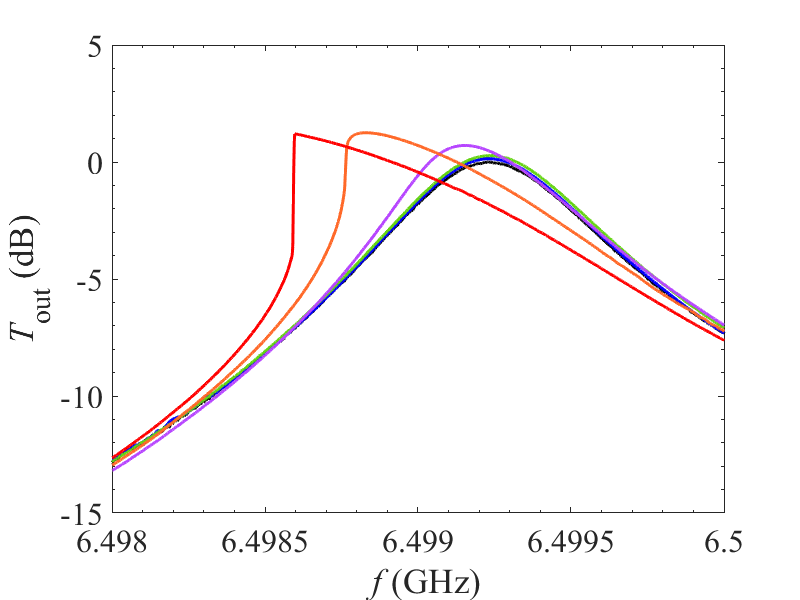}
\caption{\label{fig:Nb_Bifurcation_warm} Measurements of transmission $T_\mathrm{out}$ against frequency $f$ for a Nb half-wave resonator at $4\,\mathrm{K}$ for varying levels of sweep power $P_\mathrm{in}$. The sweep direction is from low frequency to high frequency. Black line: $P_\mathrm{in}=-20\,\mathrm{dBm}$; blue line: $P_\mathrm{in}=-10\,\mathrm{dBm}$; green line: $P_\mathrm{in}=0\,\mathrm{dBm}$; purple line: $P_\mathrm{in}=10\,\mathrm{dBm}$; orange line: $P_\mathrm{in}=15\,\mathrm{dBm}$; red line: $P_\mathrm{in}=17.5\,\mathrm{dBm}$.}
\end{figure}

The transition temperatures of our thin Nb films are around $9\,\mathrm{K}$. This allows the films to remain in their superconducting states at a temperature maintained by a pulse tube cooler, i.e. $\sim4\,\mathrm{K}$. Figure~\ref{fig:Nb_Bifurcation_warm} shows transmission measurement of the Nb half-wave resonator at $4\,\mathrm{K}$ for sweep power $P_\mathrm{in}$ from $-20\,\mathrm{dBm}$ to $17.5\,\mathrm{dBm}$. In contrast with the transmission measurement at $0.1\,\mathrm{K}$, the transmission measurement at $4\,\mathrm{K}$ no longer shows discernible increase in resonant frequency at low sweep powers. At low sweep powers of $-20\,\mathrm{dBm}$ to $0\,\mathrm{dBm}$, the resonant frequency remain unchanged and the transmission of the resonator increases slightly (by $0.3\,\mathrm{dB}$) as the sweep power is increased. This is qualitatively consistent with the behaviour of two-level-system nonlinear mechanism which can be significant at low sweep powers \citenumns{Jiansong_2012,Thomas_2020}. At high sweep powers of $0\,\mathrm{dBm}$ to $17.5\,\mathrm{dBm}$, as the sweep power increases, the resonant frequency decreases significantly while the transmission continues to increase.

\begin{figure}[!ht]
\includegraphics[width=8.6cm]{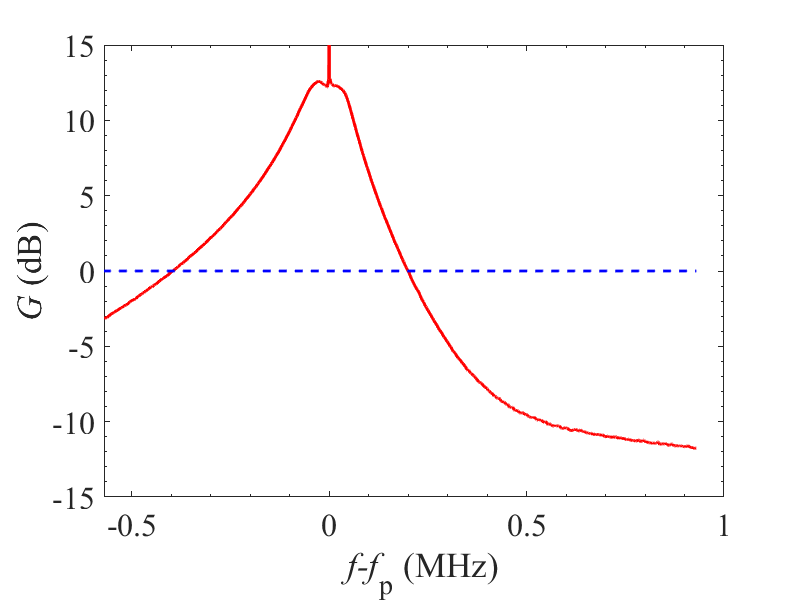}
\caption{\label{fig:Nb_amplification_warm} Measurements of power gain $G$ against frequency $f-f_\mathrm{p}$ for a Nb half-wave resonator at $4\,\mathrm{K}$. The sharp spike at $f-f_\mathrm{p}=0\,\mathrm{MHz}$ is the strong pump tone. The power of the pump tone at the input of the cryostat $P_\mathrm{pump}$ is $15\,\mathrm{dBm}$ and it is placed at frequency close to the peak of the resonance. }
\end{figure}

Figure~\ref{fig:Nb_amplification_warm} shows amplification measurement of the same Nb resonator at $4\,\mathrm{K}$ using a pump with $P_\mathrm{pump}=15\,\mathrm{dBm}$ at frequency close to the peak of the resonance. Compared to the amplification measurement at $0.1\,\mathrm{K}$, much higher pump power is required to achieve gain at $4\,\mathrm{K}$. This is because, at higher temperatures, dissipation increases and results in a decrease in the Q-factor of the resonance. A smaller Q-factor increases both the bandwidth of the resonance without a pump tone and the bandwidth of amplification in the presence of a pump tone. As seen in Figure~\ref{fig:Nb_amplification_warm}, gains of $>10\,\mathrm{dB}$ can now be achieved over an extended range of $140\,\mathrm{kHz}$ at $4\,\mathrm{K}$, i.e. a seven-fold increase over the $>10\,\mathrm{dB}$ bandwidth at $0.1\,\mathrm{K}$. In practise, the Q-factor of a resonator amplifier can be more reliably controlled by engineering the coupling capacitances at the ends of a resonator. The wide bandwidth amplification shown in Figure~\ref{fig:all_compiled} is the result of reducing the Q-factor in this manner. The fact that high gain can be achieved at $4\,\mathrm{K}$ is highly important from a practical perspective: using a pulse tube cooler to control the environment of superconducting parametric amplifiers will significantly reduce the complexity and cost of operating these devices.

\section{Niobium Nitride measurements}
Alloy or amorphous materials, such as NbTiN, NbN, and WSi, are commonly used for superconducting parametric amplifiers (based on kinetic inductances) because their high resistivity and, as a result, low power requirement to achieve significant nonlinear wave-mixing \citenumns{Eom_2012,songyuan2019_nonlinear}. In order to study the nonlinear properties of these materials, we have deposited $100\,\mathrm{nm}$ NbN resonators and characterised their behaviour using the same measurement techniques as described in earlier sections.

\begin{figure}[!ht]
\includegraphics[width=8.6cm]{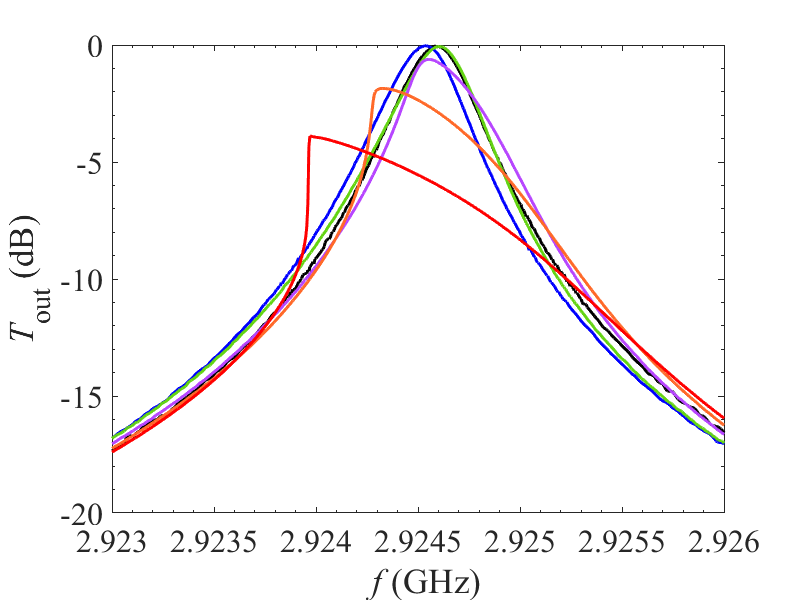}
\caption{\label{fig:NbN_Bifurcation_cold_fundamental} Measurements of transmission $T_\mathrm{out}$ against frequency $f$ for a NbN half-wave resonator at $0.1\,\mathrm{K}$ for varying levels of sweep power $P_\mathrm{in}$. The sweep direction is from low frequency to high frequency over the fundamental resonance. Black line: $P_\mathrm{in}=-70\,\mathrm{dBm}$; blue line: $P_\mathrm{in}=-60\,\mathrm{dBm}$; green line: $P_\mathrm{in}=-50\,\mathrm{dBm}$; purple line: $P_\mathrm{in}=-40\,\mathrm{dBm}$; orange line: $P_\mathrm{in}=-35\,\mathrm{dBm}$; red line: $P_\mathrm{in}=-30\,\mathrm{dBm}$.}
\end{figure}

The NbN films tested in this study have transition temperatures $T_\mathrm{c}=9.8\,\mathrm{K}$ and resistivities $\rho=1010\,\mathrm{\mu\Omega\,cm}$. Figure~\ref{fig:NbN_Bifurcation_cold_fundamental} shows transmission measurement of a NbN half-wave resonator at $0.1\,\mathrm{K}$ for varying levels of sweep power $P_\mathrm{in}$ from $-70\,\mathrm{dBm}$ to $-30\,\mathrm{dBm}$. The direction of these sweeps is from low frequency to high frequency. The resonant frequency initially decreases over the power range of $-70\,\mathrm{dBm}$ to $-60\,\mathrm{dBm}$; it then increases back to almost its original position over the power range of $-60\,\mathrm{dBm}$ to $-50\,\mathrm{dBm}$; finally the resonant frequency continuously decreases at powers greater than $-50\,\mathrm{dBm}$, before reaching bifurcation at around $-30\,\mathrm{dBm}$. Similar to the Nb, the nonlinear behaviour of NbN at low powers is complicated and is likely to be the result of several competing mechanisms. In the high power regime, which is relevant to the operation of parametric amplifiers, the behaviour of NbN is more straightforward and can be analysed using theories based on Equation~(\ref{eq:nonlinear_scale}), for example \citenumns{Thomas_2022}.

\begin{figure}[!ht]
\includegraphics[width=8.6cm]{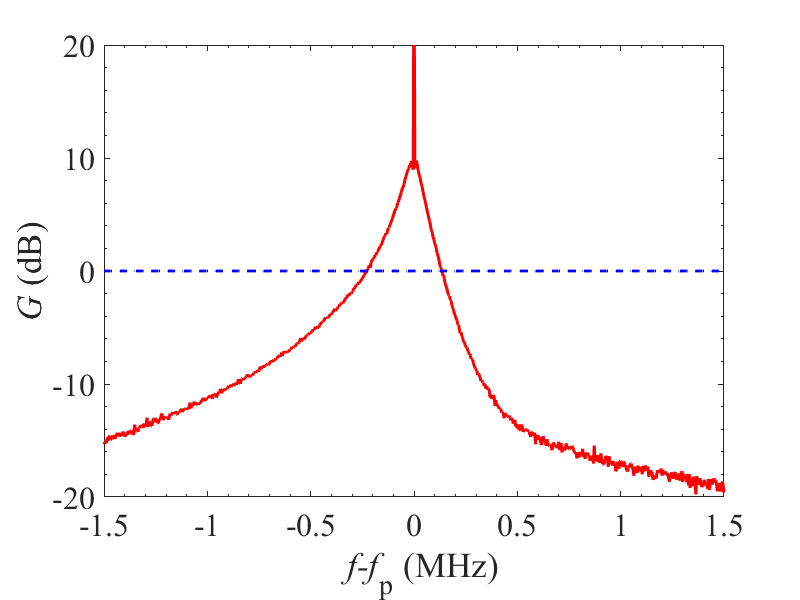}
\caption{\label{fig:NbN_amplification_cold_fundamental} Measurements of power gain $G$ against frequency $f-f_\mathrm{p}$ for a NbN half-wave resonator at $0.1\,\mathrm{K}$. The sharp spike at $f-f_\mathrm{p}=0\,\mathrm{MHz}$ is the strong pump tone. The power of the pump tone at the input of the cryostat $P_\mathrm{pump}$ is $-32\,\mathrm{dBm}$ and it is placed at frequency close to the peak of the resonance. }
\end{figure}

Figure~\ref{fig:NbN_amplification_cold_fundamental} shows amplification measurement of the same NbN resonator using a pump with $P_\mathrm{pump}=-32\,\mathrm{dBm}$ at the input of the cryostat. Moderate peak gain of $\sim10\,\mathrm{dB}$ was obtained. Similar to Nb, the amplification does not show bandwidth restriction from rate-limiting mechanism in the underlying nonlinearity. NbN has a much higher resistivity compared to Nb; consequently, a much smaller pump power is required to achieve high gain.

\begin{figure}[!ht]
\includegraphics[width=8.6cm]{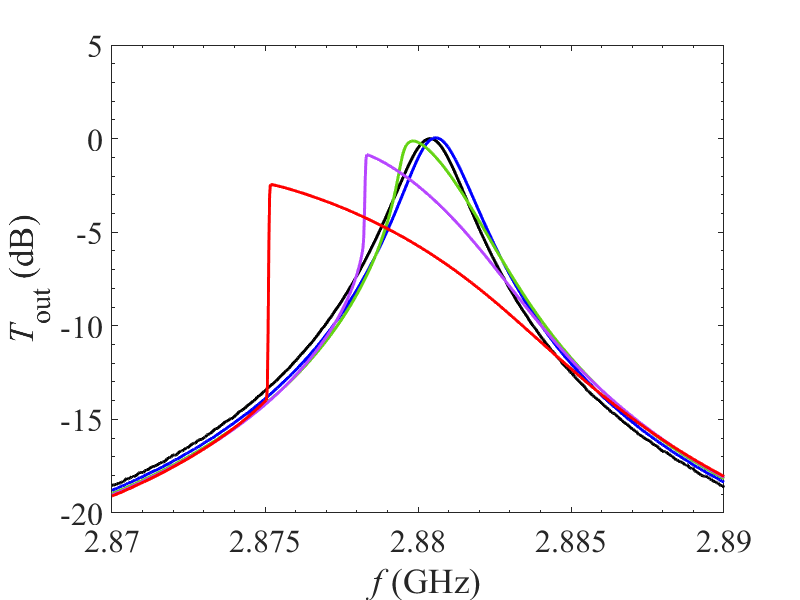}
\caption{\label{fig:NbN_Bifurcation_warm_fundamental} Measurements of transmission $T_\mathrm{out}$ against frequency $f$ for a NbN half-wave resonator at $4\,\mathrm{K}$ for varying levels of sweep power $P_\mathrm{in}$. The sweep direction is from low frequency to high frequency over the fundamental resonance. Black line: $P_\mathrm{in}=-40\,\mathrm{dBm}$; blue line: $P_\mathrm{in}=-30\,\mathrm{dBm}$; green line: $P_\mathrm{in}=-20\,\mathrm{dBm}$; purple line: $P_\mathrm{in}=-15\,\mathrm{dBm}$; red line: $P_\mathrm{in}=-10\,\mathrm{dBm}$.}
\end{figure}

Figure~\ref{fig:NbN_Bifurcation_warm_fundamental} shows transmission sweep measurements of the NbN half-wave resonator at $4\,\mathrm{K}$ for varying levels of sweep power $P_\mathrm{in}$ from $-40\,\mathrm{dBm}$ to $-10\,\mathrm{dBm}$. Over the power range of $-40\,\mathrm{dBm}$ to $-30\,\mathrm{dBm}$, the resonant frequency increases; at sweep powers higher than $-30\,\mathrm{dBm}$, the resonant frequency decreases. In the case of Nb resonators, the increase in resonant frequency occurs at $0.1\,\mathrm{K}$ at low sweep powers, but disappears at $4\,\mathrm{K}$; in contrast, in the case of NbN, this increase in resonant frequency at low sweep powers persists at $4\,\mathrm{K}$.

\begin{figure}[!ht]
\includegraphics[width=8.6cm]{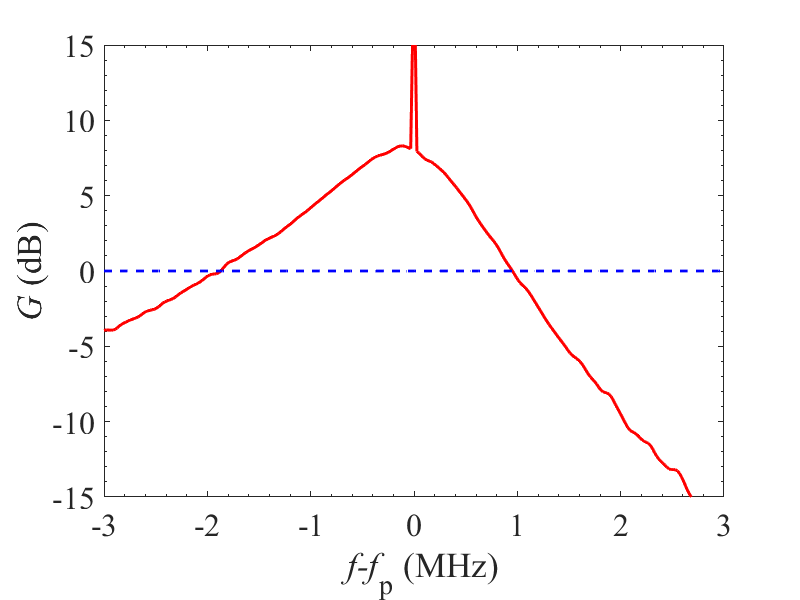}
\caption{\label{fig:NbN_amplification_warm_fundamental} Measurements of power gain $G$ against frequency $f-f_\mathrm{p}$ for a NbN half-wave resonator at $4\,\mathrm{K}$. The sharp spike at $f-f_\mathrm{p}=0\,\mathrm{MHz}$ is the strong pump tone. The power of the pump tone at the input of the cryostat $P_\mathrm{pump}$ is $-11.50\,\mathrm{dBm}$ and it is placed at frequency close to the peak of the resonance. }
\end{figure}

Figure~\ref{fig:NbN_amplification_warm_fundamental} shows amplification measurement at $4\,\mathrm{K}$ using a pump with $P_\mathrm{pump}=-11.50\,\mathrm{dBm}$ at the input of the cryostat. Similar to the trend observed in Nb resonators, much higher pump power is required to achieve gain at $4\,\mathrm{K}$ compared to $0.1\,\mathrm{K}$. This is also likely the result of a decrease in Q-factor of the resonance at higher temperatures. Correspondingly, the bandwidth across which the gain is positive is very wide at $\sim3\,\mathrm{MHz}$. When Nb is operated at pulse tube cooler temperatures, one of the main practical difficulties lies with maintaining a stable cryogenic environment in the presence of a very strong pump tone. The fact that NbN films have both high transition temperatures as well as high resistivities (and hence low pump power requirements) means that NbN could be especially suitable to operate as parametric amplifiers at $4\,\mathrm{K}$.

In general, a transmission line based half-wave resonator, such as the resonator design shown in Figure~\ref{fig:resonator_diagram}, will exhibit resonance at approximately integer harmonics of the fundamental resonant frequency. Whether the harmonic resonances display the same nonlinear behaviour as the fundamental resonance is an interesting question both theoretically as well as practically.

\begin{figure}[!ht]
\includegraphics[width=8.6cm]{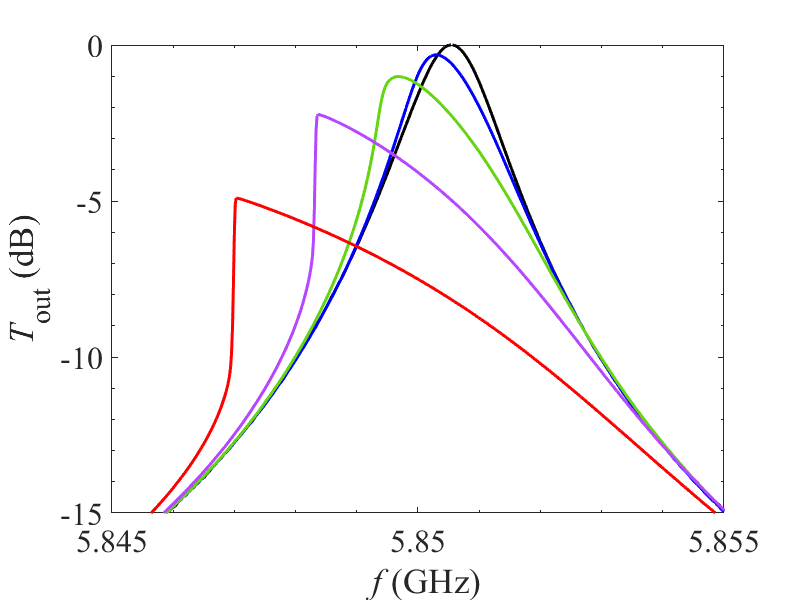}
\caption{\label{fig:NbN_Bifurcation_cold_harmonic} Measurements of transmission $T_\mathrm{out}$ against frequency $f$ for a NbN half-wave resonator at $0.1\,\mathrm{K}$ for varying levels of sweep power $P_\mathrm{in}$. The sweep direction is from low frequency to high frequency over the first harmonic. Black line: $P_\mathrm{in}=-40\,\mathrm{dBm}$; blue line: $P_\mathrm{in}=-30\,\mathrm{dBm}$; green line: $P_\mathrm{in}=-25\,\mathrm{dBm}$; purple line: $P_\mathrm{in}=-20\,\mathrm{dBm}$; red line: $P_\mathrm{in}=-15\,\mathrm{dBm}$.}
\end{figure}

Figure~\ref{fig:NbN_Bifurcation_cold_harmonic} shows transmission measurement of the same NbN half-wave resonator at $0.1\,\mathrm{K}$ around its first harmonic at $5.85\,\mathrm{GHz}$ for varying levels of sweep power $P_\mathrm{in}$ from $-40\,\mathrm{dBm}$ to $-15\,\mathrm{dBm}$. The direction of these sweeps is from low frequency to high frequency. Interestingly, and in contrast with the behaviour of NbN in the fundamental resonance, the resonant frequency of the first harmonic decreases at all applied powers. This behaviour is also seen in the other harmonics of the NbN resonator. This observation suggests that the nonlinear mechanism which is responsible for the increase of resonant frequency at low sweep powers may be suppressed in the harmonics of the NbN resonator.

\begin{figure}[!ht]
\includegraphics[width=8.6cm]{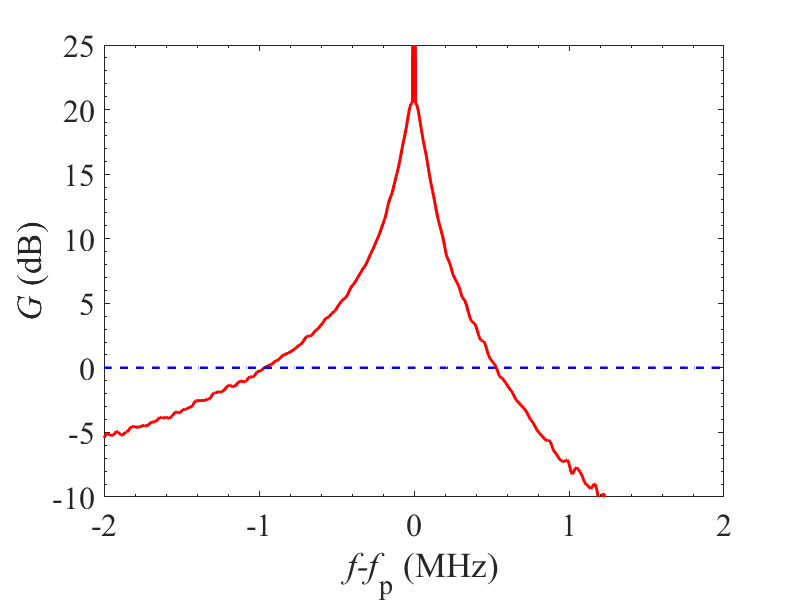}
\caption{\label{fig:NbN_amplification_cold_harmonic} Measurements of power gain $G$ against frequency $f-f_\mathrm{p}$ for a NbN half-wave resonator around its first harmonic at $0.1\,\mathrm{K}$. The sharp spike at $f-f_\mathrm{p}=0\,\mathrm{MHz}$ is the strong pump tone. The power of the pump tone at the input of the cryostat $P_\mathrm{pump}$ is $-19.20\,\mathrm{dBm}$ and it is placed at frequency close to the peak of the resonance. }
\end{figure}

Figure~\ref{fig:NbN_amplification_cold_harmonic} shows the amplification measured around the first harmonic of the resonator at $0.1\,\mathrm{K}$ using a pump with $P_\mathrm{pump}=-19.20\,\mathrm{dBm}$ at the input of the cryostat. As seen in the figure, very high peak gain of $\sim20\,\mathrm{dB}$ can be obtained. Gains higher than $10\,\mathrm{dB}$ can be achieved over a frequency width of $\sim0.5\,\mathrm{MHz}$. When using these resonator amplifiers, depending on the experimental requirement, one may choose to place the pump tone away from the optimum frequency, i.e. peak of the shifted underlying resonance, in order to widen the bandwidth at the cost of reduced gain. The pump power at the input of the parametric amplifier is around $-34.20\,\mathrm{dBm}$ which is remarkably low for the high peak gain that was achieved. Even at this early stage of development, the NbN harmonic resonator amplifier already shows many desirable characteristics such as high gain, low pump power requirement, moderate bandwidth, and the potential to operate at higher temperatures.

\section{Cross-harmonic Amplification}

\begin{figure*}
  \includegraphics[width=\textwidth]{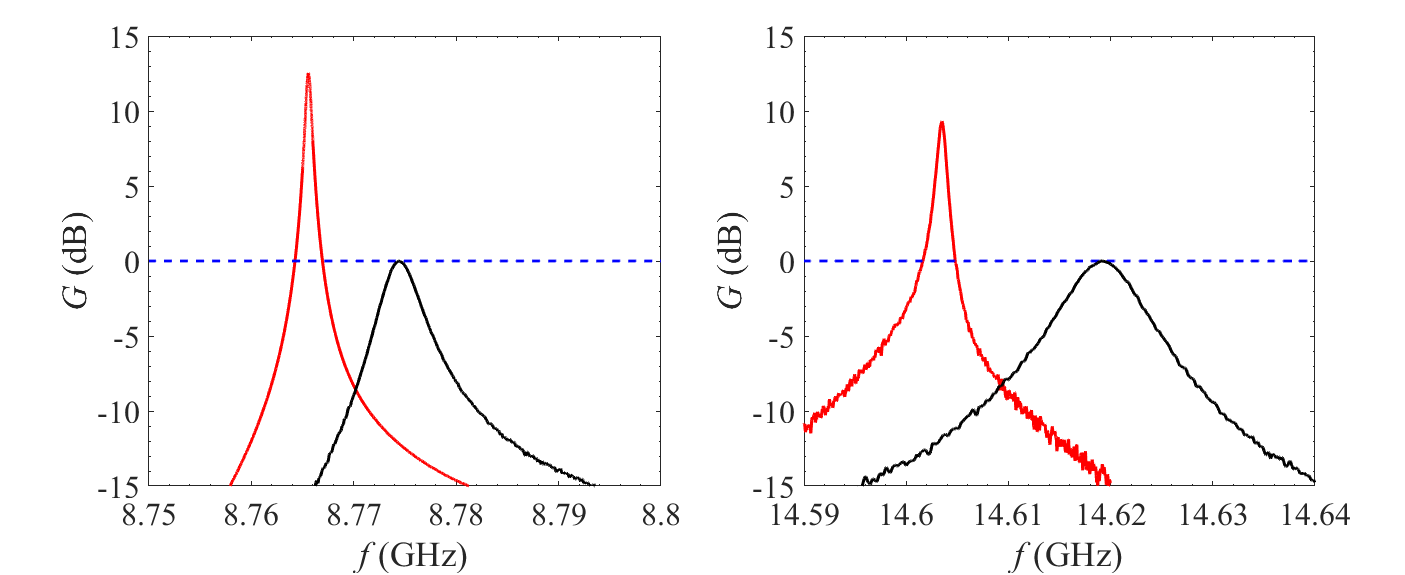}
  \caption{\label{fig:cross_amp_linear} Measurements of power gain $G$ against frequency $f-f_\mathrm{p}$ for a NbN half-wave resonator at $0.1\,\mathrm{K}$. The pump power is $3.8\,\mathrm{dBm}$ and the pump tone is placed around the peak of the resonance at $12\,\mathrm{GHz}$. Left figure shows amplification in the $9\,\mathrm{GHz}$ harmonic, whereas right figure shows amplification in the $15\,\mathrm{GHz}$ harmonic. Black line: response of the harmonics in the absence of the pump tone; red line: response of the harmonics in the presence of the pump tone. }
\end{figure*}

\begin{figure*}
  \includegraphics[width=\textwidth]{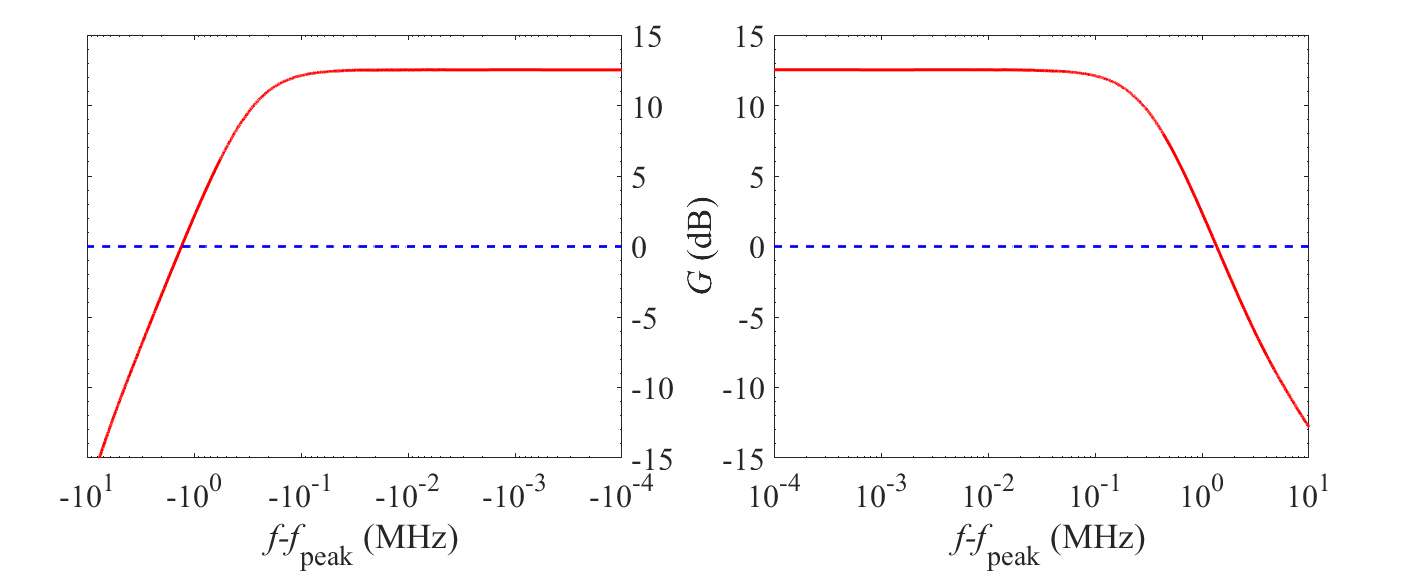}
  \caption{\label{fig:cross_amp_log} Measurements of power gain $G$ of the $9\,\mathrm{GHz}$ harmonic against frequency offset $f-f_\mathrm{peak}$ for a NbN half-wave resonator at $0.1\,\mathrm{K}$, where $f_\mathrm{peak}$ is the frequency at which the gain is maximum. The pump power is $3.8\,\mathrm{dBm}$ and the pump tone is placed around the peak of the resonance at $12\,\mathrm{GHz}$. Left figure shows the response when $f<f_\mathrm{peak}$ whereas right figure shows the response when $f>f_\mathrm{peak}$.}
\end{figure*}

The form of the nonlinearity given in Equation~(\ref{eq:nonlinear_scale}) constrains the operation of the amplifier to four-wave mixing, in the absence of a DC biasing term \citenumns{zhao2022physics}. Within this mixing regime, the signal frequency $\omega_\mathrm{s}$, idler frequency $\omega_\mathrm{i}$, and pump frequency $\omega_\mathrm{p}$ are related by $\omega_\mathrm{s}+\omega_\mathrm{i}=2\,\omega_\mathrm{p}$, i.e. the signal and idler frequencies are equally spaced around the pump frequency. There is no restriction, however, that the signal, idler, and pump tones need to be placed around the same harmonic. Since the harmonics of a half-wave resonator are approximately equally spaced, an interesting mode of operation would be to place the signal, idler, and pump tones each in separate harmonics. Figure~\ref{fig:cross_amp_linear} shows the power gain measurements of the $9\,\mathrm{GHz}$ harmonic and the $15\,\mathrm{GHz}$ harmonic in the presence of a strong pump tone in the $12\,\mathrm{GHz}$ harmonic. As seen in the figure, high gain of $\sim10\,\mathrm{dB}$ can be achieved in both harmonics. The width across which $G>0\,\mathrm{dB}$ is $\sim3\,\mathrm{MHz}$ for both harmonics. In contrast with amplification within a harmonic, the entire amplification band can be used for signal amplification as there is no need to avoid the frequencies occupied by the pump tone. 

Figure~\ref{fig:cross_amp_log} shows  the amplification in the $9\,\mathrm{GHz}$ harmonic plotted in logarithmic frequency scale around the frequency at which the gain is maximum, $f_\mathrm{peak}$. Harmonic amplification displays the typical characteristics of an amplifier: the gain shows a flat response across its bandwidth before hitting its $3\,\mathrm{dB}$ frequency corners on either side (located at $|f-f_\mathrm{peak}|=0.3\,\mathrm{MHz}$), beyond which the gain decays rapidly with $|f-f_\mathrm{peak}|$.

The fact that cross-harmonic amplification can be achieved at all strongly suggests that the underlying inductance nonlinearity of NbN has a very fast response time, at least of the order of nanoseconds. Cross-harmonic amplification is potentially an advantageous mode of operation. We are not aware of publications which has observed this mode of operation previously. It has the important benefit of allowing simpler schemes to remove the pump tone from the signal tone, for example through a low/high pass filter. This frequency separation also prevents the contamination of the signal tone by the noise from the pump tone.

\section{Conclusions}
We have performed a systematic comparison of superconducting Ti, Nb, and NbN resonators in the context of parametric amplifiers. We have successfully operated these resonators to achieve stable, high-gain amplification in accordance with theory \citenumns{Thomas_2022}. Specifically, amplifiers based on our highly resistive NbN films are able to achieve greater than $20\,\mathrm{dB}$ of power gain whilst requiring a very small pump power of $-34.20\,\mathrm{dBm}$. Our microwave measurements show that these materials exhibit a wide range of different properties and behaviours which we have interpreted using our previous theoretical analyses of nonlinear resonators and parametric amplifiers \citenumns{Thomas_2020,Thomas_2022}. Our measurements on Ti show that its inductance has a long characteristic time ($\sim74\,\mathrm{\mu s}$) when responding to a change in current, restricting the bandwidth of its frequency mixing processes. Further, Ti displays significant nonlinear dissipation which prevents the material from achieving amplification despite significant wave-mixing. Our measurements on Nb and NbN show that, at low powers, the underlying nonlinear mechanisms may \textit{decrease} the effective inductance. This \textit{decrease} in inductance is most significant at low temperatures ($\sim 100\,\mathrm{mK}$) in the fundamental resonant mode; it reduces in significance at high temperatures ($\sim 4\,\mathrm{K}$) or in the harmonic resonances.

We have shown that for a half-wave resonator, amplification can be achieved not only in the fundamental resonance but also in the higher harmonic resonances. We have further shown that for materials with high $T_\mathrm{c}$, e.g. Nb and NbN, amplification can be achieved at a temperature maintained by a pulse tube cooler, i.e. $\sim4\,\mathrm{K}$. Both observations improve the flexibility of resonator-based amplifier designs and expand their potential application. Finally, in materials systems that have very fast response times, e.g. NbN, we have found that a cross-harmonic type of amplification can be achieved by placing pump tone in a different resonant mode as the signal and the idler. This mode of operating a resonator parametric amplifier is likely to have important practical value as it allows simpler schemes to isolate and extract the signal tone from the pump tone.

\begin{acknowledgements}
The authors are grateful for funding from the UK Research and Innovation (UKRI) and the Science and Technology Facilities Council (STFC) through the Quantum Technologies for Fundamental Physics (QTFP) programme (Project Reference ST/T006307/2).
\end{acknowledgements}

\bibliographystyle{h-physrev}
\bibliography{library}

\begin{thebibliography}{10}

\bibitem{Day_2003}
P.~K. {Day}, H.~G. {LeDuc}, B.~A. {Mazin}, A.~{Vayonakis}, and J.~{Zmuidzinas},
\newblock Nature {\bf 425}, 817 (2003).

\bibitem{Dobbs_bolometer_2012}
M.~A. Dobbs {\em et~al.},
\newblock Review of Scientific Instruments {\bf 83}, 073113 (2012),
  https://doi.org/10.1063/1.4737629.

\bibitem{Mates_bolometer_2017}
J.~A.~B. Mates {\em et~al.},
\newblock Applied Physics Letters {\bf 111}, 062601 (2017),
  https://doi.org/10.1063/1.4986222.

\bibitem{Martinis2009}
J.~M. Martinis,
\newblock Quantum Information Processing {\bf 8}, 81 (2009).

\bibitem{Wesenberg_2009}
J.~H. Wesenberg {\em et~al.},
\newblock Phys. Rev. Lett. {\bf 103}, 070502 (2009).

\bibitem{Wallraff_2004}
A.~Wallraff {\em et~al.},
\newblock Nature {\bf 431}, 162 (2004).

\bibitem{Mallet_2009}
F.~Mallet {\em et~al.},
\newblock Nature Physics {\bf 5}, 791 (2009).

\bibitem{Jonas_review}
J.~Zmuidzinas,
\newblock {Annu. Rev. Condens. Matter Phys.} {\bf {3}}, 169 ({2012}).

\bibitem{Superconducting_Qubits_review}
M.~Kjaergaard {\em et~al.},
\newblock Annual Review of Condensed Matter Physics {\bf 11}, 369 (2020),
  https://doi.org/10.1146/annurev-conmatphys-031119-050605.

\bibitem{Project8_2009}
B.~Monreal and J.~A. Formaggio,
\newblock Phys. Rev. D {\bf 80}, 051301 (2009).

\bibitem{Hao_2019}
L.~Hao,
\newblock Quantum technology for cyclotron radiation detection,
\newblock Absolute Neutrino Mass Workshop at UCL, 2019.

\bibitem{Saakyan_2020}
R.~Saakyan,
\newblock Determination of neutrino mass with quantum technologies,
\newblock UK HEP Forum 2020: Quantum leaps to the dark side at Durham
  University, 2020.

\bibitem{Thomas_2020}
C.~N. Thomas, S.~Withington, Z.~Sun, T.~Skyrme, and D.~J. Goldie,
\newblock New Journal of Physics {\bf 22}, 073028 (2020).

\bibitem{Swenson_2013}
L.~J. Swenson {\em et~al.},
\newblock Journal of Applied Physics {\bf 113}, 104501 (2013),
  https://doi.org/10.1063/1.4794808.

\bibitem{Zhao_2020}
S.~Zhao, S.~Withington, D.~J. Goldie, and C.~N. Thomas,
\newblock Journal of Physics D: Applied Physics {\bf 53}, 345301 (2020).

\bibitem{Vissers_2015}
M.~R. Vissers {\em et~al.},
\newblock Applied Physics Letters {\bf 107}, 062601 (2015),
  https://doi.org/10.1063/1.4927444.

\bibitem{Adamyan_2016_resonator}
A.~A. Adamyan, S.~E. Kubatkin, and A.~V. Danilov,
\newblock Applied Physics Letters {\bf 108}, 172601 (2016),
  https://doi.org/10.1063/1.4947579.

\bibitem{Eom_2012}
B.~H. Eom, P.~K. Day, H.~G. LeDuc, and J.~Zmuidzinas,
\newblock Nat. Phys. {\bf 8}, 623 (2012).

\bibitem{Shan_2016}
W.~Shan, Y.~Sekimoto, and T.~Noguchi,
\newblock IEEE Transactions on Applied Superconductivity {\bf 26}, 1 (2016).

\bibitem{zhao2022physics}
S.~Zhao,
\newblock {\em Physics of superconducting travelling-wave parametric
  amplifiers},
\newblock PhD thesis, University of Cambridge, 2021.

\bibitem{Yurke_1988}
B.~Yurke {\em et~al.},
\newblock Phys. Rev. Lett. {\bf 60}, 764 (1988).

\bibitem{Movshovich_1990}
R.~Movshovich {\em et~al.},
\newblock Phys. Rev. Lett. {\bf 65}, 1419 (1990).

\bibitem{Tholen_2009}
E.~A. Tholén, A.~Ergül, K.~Stannigel, C.~Hutter, and D.~B. Haviland,
\newblock Physica Scripta {\bf 2009}, 014019 (2009).

\bibitem{Vissers_2016}
M.~R. Vissers {\em et~al.},
\newblock Applied Physics Letters {\bf 108}, 012601 (2016),
  https://doi.org/10.1063/1.4937922.

\bibitem{Malnou_2020}
M.~Malnou {\em et~al.},
\newblock PRX Quantum {\bf 2}, 010302 (2021).

\bibitem{Parker_squeezing_2022}
D.~J. Parker {\em et~al.},
\newblock Phys. Rev. Appl. {\bf 17}, 034064 (2022).

\bibitem{Ranzani_2018}
L.~Ranzani {\em et~al.},
\newblock Applied Physics Letters {\bf 113}, 242602 (2018),
  https://doi.org/10.1063/1.5063252.

\bibitem{Zobrist_2019}
N.~Zobrist {\em et~al.},
\newblock Applied Physics Letters {\bf 115}, 042601 (2019),
  https://doi.org/10.1063/1.5098469.

\bibitem{Vissers_2020}
M.~R. Vissers {\em et~al.},
\newblock {Demonstration of a microwave SQUID multiplexer with
  pre-amplification from a kinetic inductance traveling-wave parametric
  amplifier},
\newblock in {\em Millimeter, Submillimeter, and Far-Infrared Detectors and
  Instrumentation for Astronomy X}, edited by J.~Zmuidzinas and J.-R. Gao Vol.
  11453, International Society for Optics and Photonics, SPIE, 2020.

\bibitem{Thomas_2022}
C.~N. Thomas, S.~Withington, and S.~Zhao,
\newblock Effects of reactive, dissipative and rate-limited nonlinearity on the
  behaviour of superconducting resonator parametric amplifiers, 2022.

\bibitem{Zhao_2022}
S.~Zhao, S.~Withington, and C.~N. Thomas,
\newblock Journal of Physics D: Applied Physics {\bf 55}, 365301 (2022).

\bibitem{Adamyan_2016}
A.~A. Adamyan, S.~E. de~Graaf, S.~E. Kubatkin, and A.~V. Danilov,
\newblock Journal of Applied Physics {\bf 119}, 083901 (2016),
  https://doi.org/10.1063/1.4942362.

\bibitem{Tess_2021}
T.~K. Skyrme,
\newblock {\em Superconducting Microwave Resonators for Low Loss Sensor
  Applications},
\newblock PhD thesis, University of Cambridge, 2021.

\bibitem{Pozar_2011}
D.~Pozar,
\newblock {\em Microwave Engineering, 4th Edition} (Wiley, 2011).

\bibitem{Jurkus_1960}
A.~{Jurkus} and P.~N. {Robson},
\newblock Proceedings of the IEE - Part B: Electronic and Communication
  Engineering {\bf 107}, 119 (1960).

\bibitem{Songyuan_2019_paramp}
S.~Zhao, S.~Withington, D.~J. Goldie, and C.~N. Thomas,
\newblock Journal of Physics D: Applied Physics {\bf 52}, 415301 (2019).

\bibitem{Saturation_JPA}
L.~Planat {\em et~al.},
\newblock Phys. Rev. Appl. {\bf 11}, 034014 (2019).

\bibitem{Tholen_2007}
E.~A. Tholén {\em et~al.},
\newblock Applied Physics Letters {\bf 90}, 253509 (2007),
  https://doi.org/10.1063/1.2750520.

\bibitem{Visser_2014}
P.~J. de~Visser {\em et~al.},
\newblock Phys. Rev. Lett. {\bf 112}, 047004 (2014).

\bibitem{Jiansong_2012}
J.~Gao,
\newblock {\em The physics of superconducting microwave resonators},
\newblock PhD thesis, California Institute of Technology, 2008.

\bibitem{songyuan2019_nonlinear}
S.~Zhao, S.~Withington, D.~J. Goldie, and C.~N. Thomas,
\newblock Journal of Low Temperature Physics {\bf 199}, 34 (2020).

\end{thebibliography}
\end{document}